\documentstyle[11pt,gh2001-asp,twoside]{article}
\begin{document}
\title{Unveiling the behavior of the ionized gas in irregular galaxies}

\author{Margarita Valdez--Guti\'errez}
\affil{Instituto Nacional de Astrof\'\i sica, Optica y Electr\'onica, 
Tonantzintla, Puebla, M\'exico}
 \author{Margarita Rosado} 
\affil{Instituto de Astronom\'\i a--UNAM, M\'exico, City, M\'exico}

\section{Results and Discussion}
We present the results of a detailed kinematical and dynamical 
analysis performed in two highly interesting gas--rich Irregular 
galaxies (Irrs): IC 1613 and NGC 4449. 
The analysis has been accomplished by means of optical Fabry--Perot
interferometry mapping the H$\alpha$ and [SII] lines. 
Our interest was centered on global and local scales in both galaxies. 
The results focused on the global radial velocity field in both 
galaxies show that they are different from each other. IC 1613 
displays a highly perturbed velocity field due to local motions 
caused by superbubbles.
NGC 4449 shows a velocity field which is relatively well behaved with 
a weak decreasing range in radial velocities ($\sim$60 km s$^{-1}$) 
from NE to SW.
On local scales, we found that the HII region population for both 
galaxies displays a velocity dispersion mean value of $\sim$20 
km s$^{-1}$, which does not reflect the environment of the host 
galaxy in each case. 
The HII region population in these galaxies display standard values
according to their diameter distribution and luminosity functions 
(D$_0$= 30--65 pc; $\alpha$= 1.5--1.9).
The superbubble population in IC 1613 displays expansion velocities
(20--40 km s$^{-1}$), sizes (100--320 pc), dynamical ages 
(0.7--2 Myr) and mechanical energies (0.6--8$\times10^{50}$ ergs) 
which can be explained --in the framework of the standard model-- 
in terms of the activity derived from the massive stars (stellar 
winds and supernova explosions) at those locations.
Based on our results (Valdez--Guti\'errez et al. 2001, 2002, Rosado 
et al. 2001), we conclude that the impressive distribution and 
\hbox {kinematical/dynamical} behavior of the ionized gas in these 
two gas--rich Irrs has been moulded --on global and local scales-- 
by the activity of their massive stars; i.e. such activity has 
``etched" globally and locally the galaxies as they look today. 
Due to the nature of the present work --and as far as we know--, 
this is the first time that the kinematics and dynamics (on global 
and local scales) of the ionized gas in nearby gas--rich Irrs is 
unveiled in such detail.

\end{document}